# Title: MicroED structure of $Au_{146}$(p-MBA)$_{57}$ at subatomic resolution reveals a twinned FCC cluster.

**Authors:** Sandra Vergara[1‡], Dylan A. Lukes[2‡], Michael W. Martynowycz[3‡], Ulises Santiago[1], German Plascencia-Villa[1], Simon C. Weiss[2], M. Jason de la Cruz[3], David M. Black[1], Marcos M. Alvarez[1], Xochitl Lopez-Lozano[1], Christopher O. Barnes[2†], Guowu Lin[2], Hans-Christian Weissker[4], Robert L. Whetten[1*], Tamir Gonen[3,5*], Guillermo Calero[2*]

**Affiliations:**

[1] Department of Physics and Astronomy, The University of Texas at San Antonio, San Antonio, TX, USA.

[2] Department of Structural Biology, University of Pittsburgh, Pittsburgh, PA, USA.

[3] Howard Hughes Medical Institute, Janelia Research Campus, Ashburn, VA, USA.

[4] Aix Marseille Université, CNRS, CINaM UMR 7325, Marseille, France.

[5] Departments of Biological Chemistry and Physiology, David Geffen School of Medicine, UCLA CA, USA.

\* Correspondence to: guc9@pitt.edu, tgonen@ucla.edu, robert.whetten@utsa.edu

† Present address: Division of Biology and Biological Engineering, California Institute of Technology, Pasadena, CA, USA.

‡ These authors contributed equally to this work.

**Abstract:** Solving the atomic structure of metallic clusters is fundamental to understanding their optical, electronic, and chemical properties. We report the structure of $Au_{146}$(p-MBA)$_{57}$ at subatomic resolution (0.85 Å) using electron diffraction (MicroED) and atomic resolution by X-ray diffraction. The 146 gold atoms may be decomposed into two constituent sets consisting of 119 core and 27 peripheral atoms. The core atoms are organized in a twinned FCC structure whereas the surface gold atoms follow a $C_2$ rotational symmetry about an axis bisecting the twinning plane. The protective layer of 57 p-MBAs fully encloses the cluster and comprises bridging, monomeric, and dimeric staple motifs. $Au_{146}$(p-MBA)$_{57}$ is the largest cluster observed exhibiting a bulk-like FCC structure as well as the smallest gold particle exhibiting a stacking fault.

**One Sentence Summary:** We report the structure of $Au_{146}$(p-MBA)$_{57}$ by MicroED and X-ray diffraction, observing for the first time a twinned FCC cluster.

**Main Text:**

Within the field of nanotechnology, metal clusters and nanoparticles are of central interest for their optical, electronic, and chemical properties. However, advances in this field have been limited by a lack of fidelity in the growth of nanoparticles. One strategy to achieve precise control over the shape and size of nanoparticles is the use of ligands as surface-protecting agents. This approach has enabled the synthesis of numerous stable clusters with atomically well-defined composition. Spectroscopic studies reveal that smaller metallic clusters have molecule-like electronic structures while larger ones support collective plasmon excitation similar to bulk-like nanoparticles (*1-3*). Solving the atomic structure of protected metal clusters is fundamental to understanding their size-dependent properties. Single crystal X-ray diffraction (XRD) has previously revealed the structure of several thiolate-protected gold clusters (*3-14*). Although some small clusters display FCC-like kernels (*3-5*), for several clusters –including the larger ones with {102, 130, 133, & 246} Au atoms– non-crystalline icosahedra and truncated decahedra (known as multiply twinned particles, or MTPs) have been found (*11-14*). However, for sufficiently large clusters, a transition to bulk-like FCC packing should occur. Although some studies suggest a geometric and electronic transition occurring in the range of $Au_{144}$ to $Au_{329,}$ the critical size for this transition is not clear (*1, 2, 15-17*). Therefore, the case of ubiquitous clusters of ~29kDa core mass (1.7-nm core diameter), commonly identified as $Au_{144}(SR)_{60,}$ is of particular interest. The standard structure-model predicts that $Au_{144}(SR)_{60}$ should have the icosahedral symmetry (*18*) which correlates well with electron microscopy studies (*19*), whereas recent powder diffraction studies suggest polymorphism in aqueous-phase samples, with competing icosahedral and truncated decahedral cores (*20*). However, the precise composition and structure of the aqueous ~29kDa cluster remains elusive.

Herein we present the structure of the largest aqueous gold cluster, $Au_{146}$(p-MBA)$_{57}$ (p-MBA: *para*-mercaptobenzoic acid), solved by MicroED (*21, 22*) to subatomic resolution (0.85 Å) and by X-ray diffraction at atomic resolution (1.3 Å). This composition has not heretofore been reported for the ultra-stable ~29kDa cluster (core mass). Synthesis of the gold nanoclusters was achieved by a modified two-phase method (*23*), where clusters were precipitated by incubation in cold methanol or in cold methanol with 100mM ammonium acetate (*24*). The product was separated by native-PAGE (Fig. S1). In both cases, mass spectrometric analysis via electro spray ionization mass spectroscopy (ESI-MS) methods (supplementary text) is consistent with a predominant $Au_{146}$(p-MBA)$_{57}$ and a minor $Au_{144}$(p-MBA)$_{60}$ components, as the assigned MS peak intensities showed a ratio near 2:1 in the former sample and ~ 4:1 in the latter (ammonium acetate coprecipitation) (Fig. S2 and Table S1). The composition $Au_{146}$(p-MBA)$_{57}$ was not reported previously as the small mass-difference compared to $Au_{144}$(p-MBA)$_{60}$ (65 Da of ~37,500 Da total) led to the interpretation of the multiple peaks as adducts (*25*). Thus, $Au_{146}$(p-MBA)$_{57}$ does not constitute an artifact of crystallization. Samples without ammonium acetate formed poorly-diffracting hexagonal plate crystals similar to those previously reported (*20, 26, 27*), whereas samples co-precipitated with ammonium acetate crystallized as needles and plates in the presence of 25% polyethylene glycol 8000 and 50mM sodium potassium phosphate (Table S2). Large crystals as well as fragmented nanocrystals (measuring less than 5μm, Fig. 1A) were selected for X-ray or MicroED experiments, respectively (*24*). Remarkably, subatomic resolution was only achieved with electron diffraction via MicroED where crystals were analyzed in a frozen hydrated state (Fig. 1B and Fig. S3). A total of 146 gold atoms and 57 p-MBA ligands were identified in the electron density maps correlating with ESI-MS (Fig. 1C). The surface representation of the metal cluster shows a globular and well-ordered structure that is effectively coated by p-MBA, preventing further growth (Fig. 1D and Movie S1). The overall architecture of the cluster is illustrated stereographically by Fig. 1E. The root-mean-square deviation between MicroED and X-ray atomic positions was 0.05 Å for the cluster kernel (innermost 13 Au atoms) and 0.1 Å for the entire gold structure, confirming the equivalence of the two methods. This MicroED structure represents the highest resolution structure for such aqueous phase gold clusters and the highest resolution determined to date by any cryoEM method.

The 146 gold atoms in the $Au_{146}$(p-MBA)$_{57}$ may be decomposed into two constituent sets: 119 core atoms and 27 peripheral atoms. The core atoms are organized in a twinned FCC structure (Fig. 2A), whereas the surface gold atoms follow a $C_2$ rotational symmetry about an axis bisecting the twinning plane (Fig. 2B). While FCC structures may be described morphologically as nested cuboctahedra, twinned FCC structures may be described as nested *anti*-cuboctahedra (Fig. S4). Anti-cuboctahedra are also known as J27 under the Johnson solid classification system. The eight triangular faces of a J27 correspond to {111} planes, and the six square faces to {100} planes. The core of $Au_{146}$(p-MBA)$_{57}$ is comprised of three nested J27 shells: a first shell (J27-1) formed by the 12 immediate neighbors of the central Au-atom site, a second shell (J27-2) of 42 sites, and an incomplete third shell (J27-3) comprising 60 sites (115 in total). Two pairs of additional gold atoms located on {100} facets complete the 119-atom core. The observed truncation of 32 sites, with respect to an ideal J27-3 shell, rounds the morphology (Fig. S5), as these truncations occur at sites of low coordination-number: 12 at vertices, and 20 along edges. In particular, sites on the boundary between mirror {111} planes are not occupied. In addition, in four of the {111} facets, we observe a lattice distortion that resembles the HCP packing (Fig. S6). The occupied positions as well as the distorted positions in the third shell are symmetric across mirror planes (Fig. S7). However, non-mirror {111} planes are distinct throughout the third shell. The view along the [111] direction (Fig. 2C) shows that atoms belonging to J27-1 and J27-2 as well as the majority of atoms in J27-3 lie on the expected positions for an FCC, while most of the peripheral atoms deviate from ideal positions (Fig. 2B and Fig. S8). The view along the [1-10] direction shows the twin in the $Au_{146}$ structure. We observe a strong correlation between peaks of the radial distribution function and those of an FCC structure (Fig. S9), with a peak gold-gold distance of 2.88 Å and most atoms in the range 2.73Å–3.1Å.

Alternatively, one notes the presence of a 79-site twinned truncated octahedron (t-TO+) as a substructure of $Au_{146}$ (Fig. 2D). The superior stability of the Marks decahedron and t-TO+ was previously predicted for gold clusters as compared to the icosahedron or the untwinned TO+ (*17, 28*). The 79-atom t-TO+ substructure may be recovered by capping each of the six {100} facets of J27-2 with four gold atoms. This t-TO+ also appears as a distinguishable group in the radial distance histogram from the central atom (Fig. S10).

The protective layer of $Au_{146}(p-MBA)_{57}$ comprises 7 bridging motifs (-S-), 19 monomeric staple motifs (-S-Au-S-), and 4 dimeric staple motifs (-S-Au-S-Au-S-) arranged around the $C_2$ symmetry axis bisecting the twinning plane (Fig. 3A and Fig. 3B). Four of the bridging motifs lie on {100} facets, two connect mirror {111} facets and one is located on the screw axis. This bridging motif has been observed primarily on {100} facets (*7, 10*), but has recently been identified linking {111} facets in the structure of $Au_{246}(p-MBT)_{80}$ (Fig. S11) (*14*). The monomeric staples connect adjacent facets. We observe a broader distribution of dihedral angles in these monomeric staples as compared to those in other clusters (Fig. S12). The four dimeric staples are anchored to {111} facets, two of them directly linked to atoms in the J27-2. Dimeric staples have been observed forming a "V" shape with the sulfur atoms in almost coplanar positions (*9, 11*). In Au$_{146}$(p-MBA)$_{57}$, the three sulfur atoms (-S-Au-S-Au-S-) are not coplanar resulting in a bending of the staple. Interestingly, both characteristics of dimeric staples, the exclusive binding on {111} facets and the bending, are also present in the nonaqueous system Au$_{246}$(p-MBT)$_{80}$ (*14*) (Fig. S12). The overall distribution of the staples is presented in Fig. 3C.

Starting from the experimental coordinates and either replacing the thiolate ligands (RS-) by chloride (Cl-), or the R-group by methyl (-CH3), we have analyzed the structure by density functional theory (DFT) methods (supplementary text). The optimized (relaxed) structure thus obtained retains all connectivity of the original, while removing much of its irregularities in interatomic distances, thereby enhancing the C$_2$-symmetry feature (Fig. S13). In particular, Au-S bond lengths show a bimodal distribution, in agreement with experiment, where this double-peak structure is rather broad. The first peak is at 2.297 Å and corresponds to peripheral Au-S bonds, while the second is at 2.360 Å and corresponds to core Au-S bonds.

Gold clusters may manifest both crystalline FCC and non-crystalline structures. FCC cores have been observed in some small clusters (*3-6, 8*) and more recently, in the tetragonal-shaped Au$_{92}$ (*10*). Other clusters, including the larger ones (Au$_{102}$, Au$_{130}$, Au$_{133}$, and Au$_{246}$), display cores with five-fold symmetry (MTPs) (*7, 9, 11-14*). Figure 4 shows the six largest gold clusters solved to date. Au$_{92}$ presents an FCC core; Au$_{102}$, Au$_{130}$, and Au$_{246}$ display decahedra cores; and Au$_{133}$ has a Mackay icosahedral core. Au$_{146}$ is not only the largest solved cluster with a FCC core, but also the first containing a stacking fault. The twinned-FCC Au$_{146}$ starkly contrasts with previous icosahedral and decahedral models proposed for the ubiquitous 29-kDa clusters (*18-20*).

Although icosahedral and decahedral structures display energetically-favorable close-packed outer facets, as the size of the particle increases, a preference for bulk-like FCC structures is expected. While nucleation mechanisms remain unclear, most likely FCC nanoparticles should grow starting from FCC seeds. The favorability of a cluster as a seed for an FCC particle is linked to the ability to grow quickly and leave the subcritical size region before disintegration (*29*). The twinned-FCC anti-cuboctahedral kernel of $Au_{146}$ may therefore be a more suitable seed than a cuboctahedral kernel, as the saturation of {100} facets in J27 generates extra sites with 4-coordination on the twin plane than are absent in the untwinned structure (Fig. S14). This creation of 4-coordination sites is similar to the well-established effect of twins in the growth of nanoparticles (*30*). In the case of nanoparticles, the introduction of twins favors a kinetic growth on the twin that leads to thermodynamically-unfavorable shapes like triangular plates or rods.

In this report we have shown that even a FCC structure with such few atoms (Au146, ~1.7 nm) contains a well-defined twinning plane. Considering that decahedral structures have 5 twin-planes and icosahedra have 30 twin-planes, twinning appears to be a critical component of nanoparticle structure from a very early stage of growth. Moreover, to our knowledge, the subatomic resolution structure of $Au_{146}(p-MBA)_{57}$ presented here is the first such structure of a metal cluster obtained by electron diffraction of frozen hydrated samples, establishing MicroED as a new tool for the characterization of nanomaterials.

**References and Notes:**

1. Y. Negishi, T. Nakazaki, S. Malola, S. Takano, Y. Niihori, W. Kurashige, S. Yamazoe, T. Tsukuda, H. Hakkinen, A critical size for emergence of nonbulk electronic and geometric structures in dodecanethiolate-protected Au clusters. *J. Am. Chem. Soc.* **137**, 1206-1212 (2015).

2. M. Zhou, C. J. Zeng, Y. X. Chen, S. Zhao, M. Y. Sfeir, M. Z. Zhu, R. C. Jin, Evolution from the plasmon to exciton state in ligand-protected atomically precise gold nanoparticles. *Nat. Commun.* **7**, (2016).

3. C. Zeng, Y. Chen, K. Iida, K. Nobusada, K. Kirschbaum, K. J. Lambright, R. Jin, Gold Quantum Boxes: On the Periodicities and the Quantum Confinement in the Au(2)(8), Au(3)(6), Au(4)(4), and Au(5)(2) Magic Series. *J. Am. Chem. Soc.* **138**, 3950-3953 (2016).

4. C. Zeng, T. Li, A. Das, N. L. Rosi, R. Jin, Chiral structure of thiolate-protected 28-gold-atom nanocluster determined by X-ray crystallography. *J. Am. Chem. Soc.* **135**, 10011-10013 (2013).

5. C. Zeng, H. Qian, T. Li, G. Li, N. L. Rosi, B. Yoon, R. N. Barnett, R. L. Whetten, U. Landman, R. Jin, Total Structure and Electronic Properties of the Gold Nanocrystal Au36(SR)24. *Angew. Chem. Int. Ed.* **51**, 13114-13118 (2012).

6. D. Crasto, S. Malola, G. Brosofsky, A. Dass, H. Hakkinen, Single crystal XRD structure and theoretical analysis of the chiral Au30S(S-t-Bu)18 cluster. *J. Am. Chem. Soc.* **136**, 5000-5005 (2014).

7. H. Qian, W. T. Eckenhoff, Y. Zhu, T. Pintauer, R. Jin, Total structure determination of thiolate-protected Au38 nanoparticles. *J. Am. Chem. Soc.* **132**, 8280-8281 (2010).

8. S. Yang, J. Chai, Y. Song, X. Kang, H. Sheng, H. Chong, M. Zhu, A New Crystal Structure of Au36 with a Au14 Kernel Cocapped by Thiolate and Chloride. *J. Am. Chem. Soc.* **137**, 10033-10035 (2015).

9. M. W. Heaven, A. Dass, P. S. White, K. M. Holt, R. W. Murray, Crystal structure of the gold nanoparticle [N(C8H17)4][Au25(SCH2CH2Ph)18]. *J. Am. Chem. Soc.* **130**, 3754-3755 (2008).

10. C. Zeng, C. Liu, Y. Chen, N. L. Rosi, R. Jin, Atomic Structure of Self-Assembled Monolayer of Thiolates on a Tetragonal Au92 Nanocrystal. *J. Am. Chem. Soc.* **138**, 8710-8713 (2016).

11. P. D. Jadzinsky, G. Calero, C. J. Ackerson, D. A. Bushnell, R. D. Kornberg, Structure of a thiol monolayer-protected gold nanoparticle at 1.1 A resolution. *Science* **318**, 430-433 (2007).

12. Y. Chen, C. Zeng, C. Liu, K. Kirschbaum, C. Gayathri, R. R. Gil, N. L. Rosi, R. Jin, Crystal Structure of Barrel-Shaped Chiral Au130(p-MBT)50 Nanocluster. *J. Am. Chem. Soc.* **137**, 10076-10079 (2015).

13. C. Zeng, Y. Chen, K. Kirschbaum, K. Appavoo, M. Y. Sfeir, R. Jin, Structural patterns at all scales in a nonmetallic chiral Au(133)(SR)(52) nanoparticle. *Sci. Adv.* **1**, e1500045 (2015).

14. C. Zeng, Y. Chen, K. Kirschbaum, K. J. Lambright, R. Jin, Emergence of hierarchical structural complexities in nanoparticles and their assembly. *Science* **354**, 1580-1584 (2016).

15. R. Philip, P. Chantharasupawong, H. F. Qian, R. C. Jin, J. Thomas, Evolution of Nonlinear Optical Properties: From Gold Atomic Clusters to Plasmonic Nanocrystals. *Nano Lett.* **12**, 4661-4667 (2012).

16. C. Yi, H. Zheng, L. M. Tvedte, C. J. Ackerson, K. L. Knappenberger, Nanometals: Identifying the Onset of Metallic Relaxation Dynamics in Monolayer-Protected Gold Clusters Using Femtosecond Spectroscopy. *J. Phys. Chem. C* **119**, 6307-6313 (2015).

17. C. L. Cleveland, U. Landman, T. G. Schaaff, M. N. Shafigullin, P. W. Stephens, R. L. Whetten, Structural evolution of smaller gold nanocrystals: The truncated decahedral motif. *Phys. Rev. Lett.* **79**, 1873-1876 (1997).

18. O. Lopez-Acevedo, J. Akola, R. L. Whetten, H. Grönbeck, H. Häkkinen, Structure and bonding in the ubiquitous icosahedral metallic gold cluster Au 144(SR) 60. *J. Phys. Chem. C.* **113**, 5035-5038 (2009).

19. D. Bahena, N. Bhattarai, U. Santiago, A. Tlahuice, A. Ponce, S. B. H. Bach, B. Yoon, R. L. Whetten, U. Landman, M. Jose-Yacaman, STEM electron diffraction and high-resolution images used in the determination of the crystal structure of the Au144(SR)60 cluster. *J. Phys. Chem. Lett.* **4**, 975-981 (2013).

20. K. M. Jensen, P. Juhas, M. A. Tofanelli, C. L. Heinecke, G. Vaughan, C. J. Ackerson, S. J. Billinge, Polymorphism in magic-sized Au144(SR)60 clusters. *Nat. Commun.* **7**, 11859 (2016).

21. D. Shi, B. L. Nannenga, M. G. Iadanza, T. Gonen, Three-dimensional electron crystallography of protein microcrystals. *Elife* **2**, e01345 (2013).

22. B. L. Nannenga, D. Shi, A. G. Leslie, T. Gonen, High-resolution structure determination by continuous-rotation data collection in MicroED. *Nat. Methods* **11**, 927-930 (2014).

23. G. Plascencia-Villa, B. Demeler, R. L. Whetten, W. P. Griffith, M. Alvarez, D. M. Black, M. Jose-Yacaman, Analytical Characterization of Size-Dependent Properties of Larger Aqueous Gold Nanoclusters. *J. Phys. Chem. C.* **120**, 8950-8958 (2016).

24. Materials and Methods are available as supplementary materials at the Science website.

25. M. M. Alvarez, J. Chen, G. Plascencia-Villa, D. M. Black, W. P. Griffith, I. L. Garzon, M. Jose-Yacaman, B. Demeler, R. L. Whetten, Hidden Components in Aqueous "Gold-144" Fractionated by PAGE: High-Resolution Orbitrap ESI-MS Identifies the Gold-102 and Higher All-Aromatic Au-pMBA Cluster Compounds. *J. Phys. Chem. B* **120**, 6430-6438 (2016).

26. S. Mustalahti, P. Myllyperkiö, T. Lahtinen, K. Salorinne, S. Malola, J. Koivisto, H. Häkkinen, M. Pettersson, Ultrafast Electronic Relaxation and Vibrational Cooling Dynamics of Au144(SC2H4Ph)60 Nanocluster Probed by Transient Mid-IR Spectroscopy. *J. Phys. Chem. C* **118**, 18233-18239 (2014).

27. O. A. Wong, C. L. Heinecke, A. R. Simone, R. L. Whetten, C. J. Ackerson, Ligand symmetry-equivalence on thiolate protected gold nanoclusters determined by NMR spectroscopy. *Nanoscale* **4**, 4099-4102 (2012).

28. J. P. K. Doye, D. J. Wales, Structural consequences of the range of the interatomic potential - A menagerie of clusters. *J. Chem. Soc. Faraday T.* **93**, 4233-4243 (1997).

29. B. W. Vandewaal, An Atomic-Scale Model of Fcc Crystal-Growth. *Z. Phys. D Atom. Mol. Cl.* **20**, 349-352 (1991).

30. C. Lofton, W. Sigmund, Mechanisms controlling crystal habits of gold and silver colloids. *Adv. Funct. Mater.* **15**, 1197-1208 (2005).

31. L. M. Tvedte, C. J. Ackerson, Size-Focusing Synthesis of Gold Nanoclusters with p-Mercaptobenzoic Acid. *The Journal of Physical Chemistry A* **118**, 8124-8128 (2014).

32. C. L. Heinecke, C. J. Ackerson, in *Nanoimaging: Methods and Protocols,* A. A. Sousa, M. J. Kruhlak, Eds. (Humana Press, Totowa, NJ, 2013), pp. 293-311.

33. H. P. Stevenson, G. Lin, C. O. Barnes, I. Sutkeviciute, T. Krzysiak, S. C. Weiss, S. Reynolds, Y. Wu, V. Nagarajan, A. M. Makhov, R. Lawrence, E. Lamm, L. Clark, T. J. Gardella, B. G. Hogue, C. M. Ogata, J. Ahn, A. M. Gronenborn, J. F. Conway, J. P. Vilardaga, A. E. Cohen, G. Calero, Transmission electron microscopy for the evaluation and optimization of crystal growth. *Acta. Crystallogr. D Struct. Biol.* **72**, 603-615 (2016).

34. C. O. Barnes, E. G. Kovaleva, X. Fu, H. P. Stevenson, A. S. Brewster, D. P. DePonte, E. L. Baxter, A. E. Cohen, G. Calero, Assessment of microcrystal quality by transmission electron microscopy for efficient serial femtosecond crystallography. *Arch. Biochem. Biophys.* **602**, 61-68 (2016).

35. J. Hattne, D. Shi, M. J. de la Cruz, F. E. Reyes, T. Gonen, Modeling truncated pixel values of faint reflections in MicroED images. *J. Appl. Crystallogr.* **49**, 1029-1034 (2016).

36. J. Hattne, F. E. Reyes, B. L. Nannenga, D. Shi, M. J. de la Cruz, A. G. Leslie, T. Gonen, MicroED data collection and processing. *Acta Crystallogr. A Found. Adv.* **71**, 353-360 (2015).

37. W. Kabsch, Xds. *Acta Crystallogr. D Biol. Crystallogr.* **66**, 125-132 (2010).

38. G. M. Sheldrick, SHELXT - Integrated space-group and crystal-structure determination. *Acta Crystallogr. A Found. Adv.* **71**, 3-8 (2015).

39. R. Miller, S. M. Gallo, H. G. Khalak, C. M. Weeks, Snb - Crystal-Structure Determination Via Shake-and-Bake. *J. Appl. Crystallogr.* **27**, 613-621 (1994).

40. G. M. Sheldrick, A short history of SHELX. *Acta Crystallogr A* **64**, 112-122 (2008).

41. P. D. Adams, P. V. Afonine, G. Bunkoczi, V. B. Chen, I. W. Davis, N. Echols, J. J. Headd, L. W. Hung, G. J. Kapral, R. W. Grosse-Kunstleve, A. J. McCoy, N. W. Moriarty, R. Oeffner, R. J. Read, D. C. Richardson, J. S. Richardson, T. C. Terwilliger, P. H. Zwart, PHENIX: a comprehensive Python-based system for macromolecular structure solution. *Acta Crystallogr. D Biol. Crystallogr.* **66**, 213-221 (2010).

42. D. M. Black, S. B. Bach, R. L. Whetten, Capillary Liquid Chromatography Mass Spectrometry Analysis of Intact Monolayer-Protected Gold Clusters in Complex Mixtures. *Anal. Chem.* **88**, 5631-5636 (2016).

43. D. M. Black, M. M. Alvarez, F. Yan, W. P. Griffith, G. Plascencia-Villa, S. B. H. Bach, R. L. Whetten, Triethylamine Solution for the Intractability of Aqueous Gold–Thiolate Cluster Anions: How Ion Pairing Enhances ESI-MS and HPLC of aq-Aun(pMBA)p. *J. Phys. Chem. C*, (2016).

44. D. E. Jiang, M. Walter, The halogen analogs of thiolated gold nanoclusters. *Nanoscale* **4**, 4234-4239 (2012).

45. G. Kresse, J. Furthmüller, Efficiency of ab-initio total energy calculations for metals and semiconductors using a plane-wave basis set. *Comp. Mater. Sci.* **6**, 15-50 (1996).

46. G. Kresse, D. Joubert, From ultrasoft pseudopotentials to the projector augmented-wave method. *Phys. Rev. B* **59**, 1758-1775 (1999).

47. P. Emsley, K. Cowtan, Coot: model-building tools for molecular graphics. *Acta Crystallogr. D Biol. Crystallogr.* **60**, 2126-2132 (2004).

**Acknowledgments:**

We would like to acknowledge May Elizabeth Sharp and Meitan Wang at the Swiss Light Source (SLS). We would also like to acknowledge Michael Becker and Craig Ogata at GM/CA (Argonne National Laboratory). We would also like to acknowledge John Chrzas and Rod Salazar at the South East Regional Collaborative Access Team (SERCAT, Argonne National Laboratory). The authors acknowledge Aina Cohen and Jinhu Song, from the Macromolecular Femtosecond Crystallography (MFX) at the Linac Coherent Light Source (LCLS) for their support during data collection. We thank D. Lee for home-source X-ray technical support at the University of Pittsburgh. Research reported in this publication was supported by the National Institute on Minority Health and Health Disparities (NIMHD) of the National Institutes of Health (#G12-MD007591) and by The Welch Foundation grants #AX-1615 and #AX-1857. GC acknowledges University of Pittsburgh startup funds and support from NIH grants P50-GM082251, R01-GM112686 and BioXFEL-STC1231306. GM/CA@aps has been funded in whole or in part with Federal funds from the National Cancer Institute (ACB-12002) and the National Institute of General Medicical Sciences (AGM-12006). Use of the Advanced Photon Source was supported by the U. S. Department of Energy, Office of Science, Office of Basic Energy Sciences, under Contract No. W-31-109-Eng-38. Use of the Linac Coherent Light Source (LCLS), SLAC National Accelerator Laboratory, is supported by the U.S. Department of Energy, Office of Science, Office of Basic Energy Sciences under Contract No. DE-AC02-76SF00515. The Gonen laboratory is supported by the Howard Hughes Medical Institute and this work was also supported by the Janelia Research Campus visitor program. The contents of this publication are solely the responsibility of the authors and do not necessarily represent the official views of NIGMS or NIH.

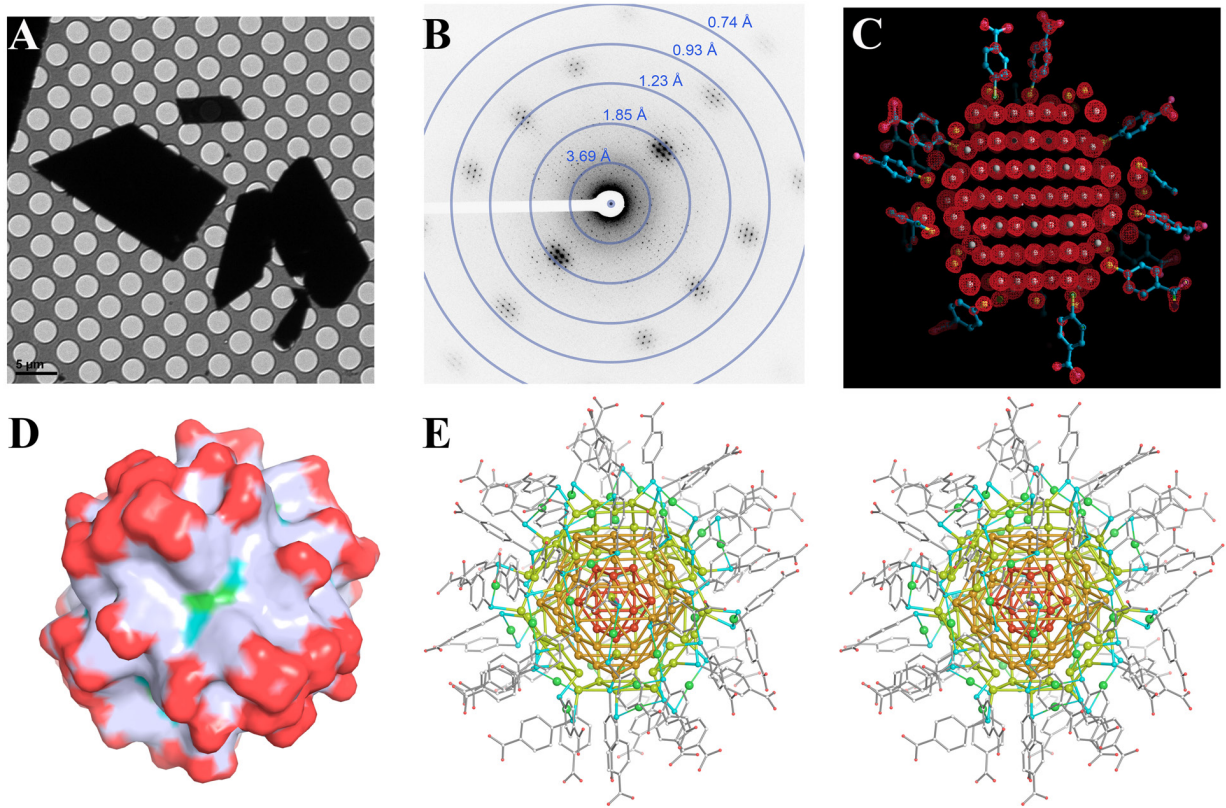

**Fig. 1**. MicroED and X-ray determination of the crystal structure of $Au_{146}$(p-MBA)$_{57}$. (**A**) Transmission electron micrograph of $Au_{146}$(p-MBA)$_{57}$ crystals and (**B**) typical MicroED data extending well beyond 1Å. (**C**) Electron diffraction density map (2Fo-Fc, contoured at 2 σ) shown as red mesh, identifies atomic positions of Au (white spheres) and S (yellow spheres) atoms. Ligands (p-MBA) are shown as blue framework. (**D**) Surface representation of the cluster with oxygen in red, carbon in white, sulfur in cyan, and gold in green. The full set of p-MBA ligands was determined using X-ray data. (**E**) Stereoscopic representation of the cluster including all gold atoms and p-MBA ligands. Gold atoms colored by shells for visualization purposes (see Fig. 2).

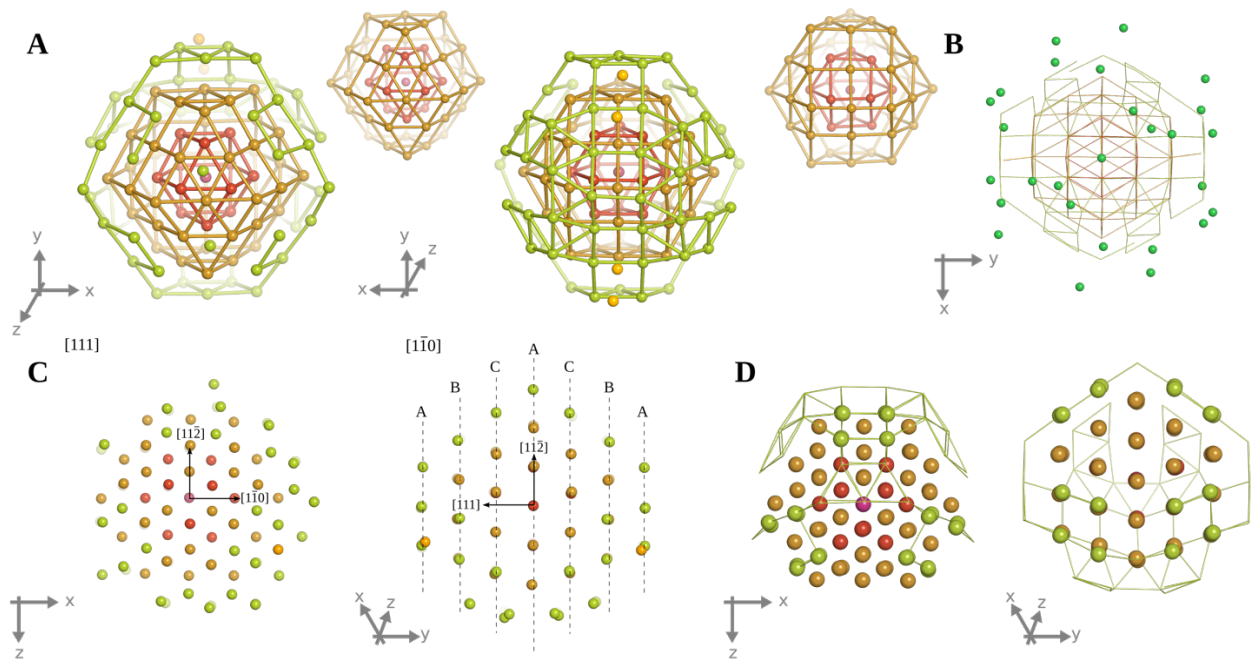

**Fig. 2.** Configuration of gold atoms in $Au_{146}$(p-MBA)$_{57}$. (**A**) Core structure (119-atoms) of the $Au_{146}$ cluster. The central atom, purple, and the first shell J27-1, red, comprise the kernel of the cluster; second shell J27-2, brown; third shell J27-3, light green; extra atoms in the core, dark yellow. Twin plane parallel to xz-plane. (**B**) Location of peripheral gold atoms (27-atoms) in dark green, with core displayed in wires. (**C**) View of crystallographic planes. Plane directions are absolute to an FCC unit cell. (**D**) Selection of 79-atom corresponding to a twinned truncated octahedron, t-TO+. Remaining atoms from core displayed on wires.

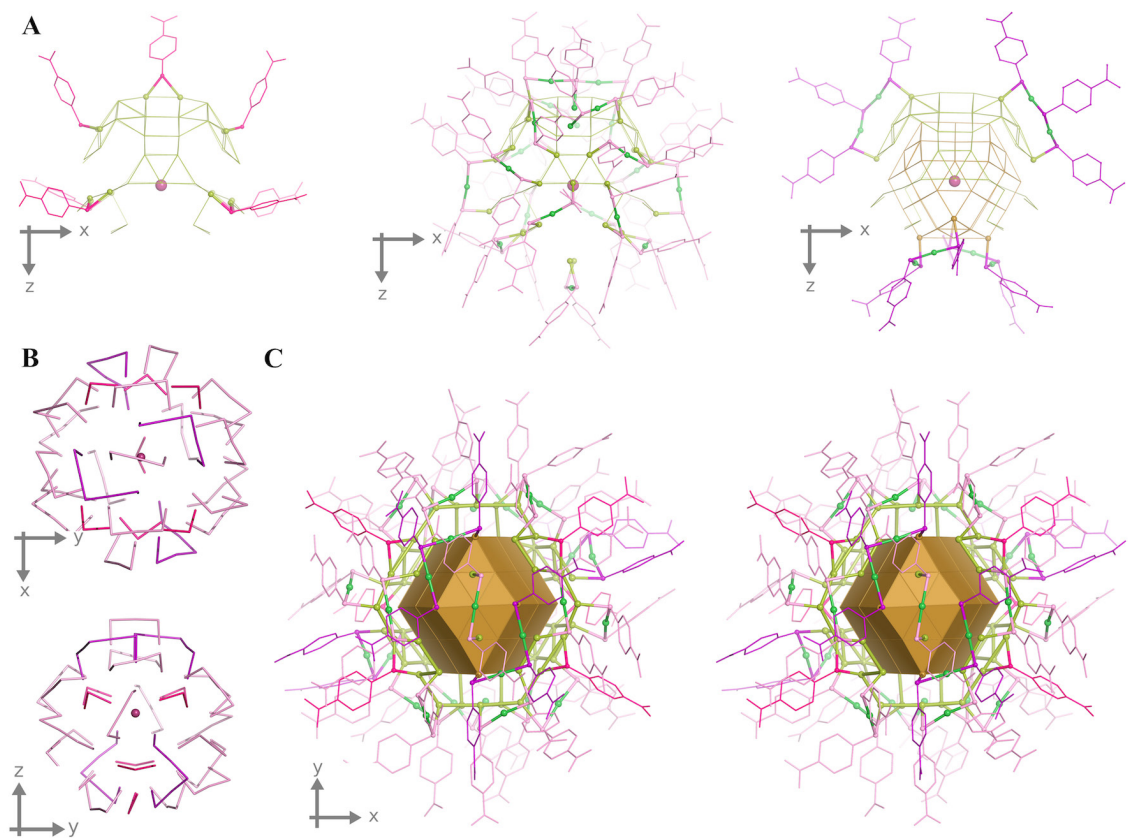

**Fig. 3**. Configuration of ligands (p-MBA) on $Au_{146}$(p-MBA)$_{57}$. (**A**) Distribution of the three types of staples on the core surface of $Au_{146}$(p-MBA)$_{57}$. Bridging motifs (left) in pink, monomeric staples (center) in light pink, and dimeric staples (right) in purple. (**B**) Rotational symmetric distribution of staples. Screw axis parallel to z-axis bisecting the twin plane (xz-plane). (**C**) Stereoscopic view of $Au_{146}$(p-MBA)$_{57}$ displaying all different types of staples, J27-2 atoms are displayed as solid surface.

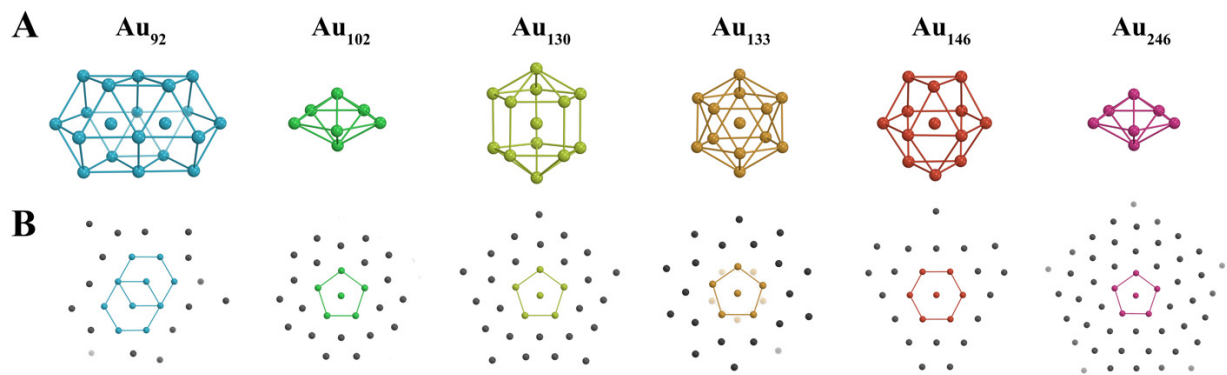

**Fig. 4.** Atomic structures of the largest gold clusters solved to date. (**A**) Kernel structures of Au$_{92}$ (conjoined cuboctahedra with no unique central atom); Au$_{102}$ and Au$_{246}$ (decahedra); Au$_{130}$ (Ino decahedron); Au$_{133}$ (icosahedron); and Au$_{146}$ (anti-cuboctahedron). (**B**) The view along the [111] axis for Au$_{92}$ (*10*) and Au$_{146}$ shows the atomic positions following the close-packing of FCC structures, whereas the projection along [110] axis for Au$_{102}$ (*11*), Au$_{130}$ (*12*), Au$_{133}$ (*13*), and Au$_{246}$ (*14*) shows five-fold symmetry.

**Supplementary Materials:**

Materials and Methods

Supplementary Text

Figures S1-S14

Tables S1-S2

Movies S1

References 31-47

# Supplementary Materials for

## MicroED Structure of Au$_{146}$(p-MBA)$_{57}$ at Subatomic Resolution Reveals a Twinned FCC Cluster.

Sandra Vergara, Dylan A. Lukes, Michael W. Martynowycz, Ulises Santiago, German Plascencia-Villa, Simon C. Weiss, , M. Jason de la Cruz, David M. Black, Marcos M. Alvarez, Xochitl Lopez-Lozano, Christopher O. Barnes, Guowu Lin, Hans-Christian Weissker, Robert L. Whetten\*, Tamir Gonen\*, Guillermo Calero\*

\*Correspondence: guc9@pitt.edu, tgonen@ucla.edu, Robert.whetten@utsa.edu

**This PDF file includes:**

- Materials and Methods
- Supplementary text
- Figs. S1 to S14
- Tables S1 and S2
- Caption for Movie S1

**Other Supplementary Materials for this manuscript includes the following:**

- Movie S1

**Materials and Methods**

Synthesis and Purification of Gold Nanoclusters
    The gold nanoclusters were produced by two-phase method with some modifications (*23, 25, 31, 32*). Stock solution of *para*-mercaptobenzoic acid (p-MBA) at 100 mM was prepared in 300 mM NaOH by vigorous stirring for at least 3 h or overnight and pulsed sonication. Nanoclusters were produced in 25 ml batch in 50 % methanol in a 50 ml round flask by adding $HAuCl_4$ to a final concentration of 3 mM and p-MBA to 9 mM. The color of this solution was light yellow. The mixture was stirred overnight until solution was colorless. Then freshly prepared ice-cold $NaBH_4$ was added to achieve 1.5 mM final concentration, and the reaction proceeded for 2 h. The entire content was transferred to a 50 ml conical tube. Volume was completed to 50 ml with cold methanol. Two distinct methods were used to precipitate the as-prepared cluster product: cold methanol or cold methanol with 100 mM ammonium acetate. The content was mixed gently and incubated at 4°C. Precipitate was concentrated with centrifugation 2000-3000 rpm 15 min, supernatant removed, and the pellet resuspended in 25 ml methanol. Nanoclusters were again concentrated by centrifugation at 2000-3000 rpm for 15 min, supernatant removed, and the pellet dried. Finally, the pellet was dissolved in pure $ddH_2O$ (0.5-1 ml). Product quality was assessed by native 10-12 % PAGE in TBE buffer at 100 V. Major bands corresponding to gold nanoclusters were cut from gel to separate fraction of interest and reconcentrated by precipitation or rotary evaporator (Fig. S1). Aliquots were stored at 4°C for further analysis. Stability of gold nanoclusters in both samples, with and without ammonium acetate coprecipitation, was monitored by routine 10-12 % PAGE and Transmission Electron Microscopy (TEM), confirming no perceivable changes over 6 months.

Gold cluster crystallization
    The two cluster samples, with and without ammonium acetate coprecipitation, were submitted to crystallization screenings. The sample without ammonium acetate formed poorly-diffracting hexagonal plates, whereas the sample coprecipitated with ammonium acetate crystallized in the presence of 25% polyethylene glycol 8000 (PEG8K) and 50 mM sodium potassium phosphate. Crystals were transferred to a stabilizing solution for cryo-protection containing 25% methyl pentanediol (MPD), 5% PEG 8K and 50mM sodium potassium phosphate.

Micro Electron Diffraction (microED).
    Crystal fragmentation: Crystals were "too thick" for observation of a crystal lattice thus a fragmentation protocol was designed. Crystals in stabilizing solution (see above) were harvested in a 0.5 ml Eppendorf tube and 30 mgs of stainless steel beads (1 mm diameter) were placed for fragmentation (*33, 34*). Sample was vortexed several times and the presence of homogeneously sized fragments was verified using bright field microscopy. Stainless steel beads were removed using a magnet and samples were centrifuged at 1000 rpms to pellet the fragments. Excess of mother liquour was removed to obtain a concentrated slurry and stabilizing cryo-protecting solution was added and mixed carefully with the slurry. This procedure was repeated three times to ensure full buffer

exchange. Holey 300 mesh carbon Quantifoil® R2/2 grids were negatively glow discharged at 25 mA and 2*10^-1 mbar for 30s from both sides. Then 2 µl of the concentrated and cryoprotected nanocrystal solution were pipetted onto the grid and incubated for 1 minute, before the excess liquid was removed by careful blotting from the backside using Fisherbrand P2 filter paper. Afterwards the grid was flash frozen by plunging it into liquid nitrogen. The grids were stored in liquid nitrogen until they were used for data collection.

Collection of MicroED Data: diffraction patterns were collected using an FEI Titan Krios TEM equipped with a field emission gun operating at 300kV, corresponding to an electron wavelength of 0.0197Å (*21, 22*). Data was collected on a TVIPS TemCam-F416 4K x 4K CMOS camera with sensor pixel dimensions of 15.6 µm x 15.6 µm. Images were taken in rolling-shutter mode with 2X-pixel binning, resulting in final images of 2048 px x 2048 px. Frames were acquired under a continuous rotation of $0.1^{o}$/s every 5 s, or a $0.5^{o}$ wedge of reciprocal space per frame. Data was converted using open-source software and corrected for pixel truncation as previously described (*35, 36*). MicroED experiments were performed at liquid nitrogen temperatures, ~77K. Nano crystals diffracted to 0.7Å resolution by MicroED.

Structure Refinement

X-ray and microED data were processed in space group P1 using the software package XDS (*37*). Reflections were indexed using the program XPREP in P2(1)/n or P2(1)/c. The X-ray structure was solved using the program SHELXT (*38*) yielding 146 gold and 57 sulfur atoms. The Fo-Fc difference map using X-ray data set 2 (Table S2) showed positive density for fourty-four pMBA molecules. The full model was completed using the difference map (Fo-Fc) from X-ray data set 1 (Table S2) which revealed all fifty-seven pMBA molecules (Fig. S3 C-E). The microED structure was solved using a partial solution obtained with the program Shake and Bake (*39*) where 79 Au atoms were clearly identified in the Patterson map. The remaining Au atoms and all 57 sulfurs were easily identified as positive density in the Fo-Fc map. Only a few p-MBA molecules were observed in the microED data.

The microED and X-ray structures were refined using the programs SHELXL (*40*) and Phenix (*41*). Several rounds of manual building and refinement were employed to place the pMBA molecules.

**Supplementary Text**

Mass Spectrometry

We previously reported a detailed electrospray ionization mass spectrometric (ESI-MS) analysis of the cluster sample without ammonium acetate coprecipitation (*25*). In that analysis, the dominant species was reported as (144,60) while stressing the presence of other "hidden" components, including intact (130,50), and ESI fragmentation products of (137,56), (102,44), and (180,66). In the light of the new composition found for the ubiquitous ~29kDa cluster (core mass), (146,57), the original mass spectra of samples without ammonium acetate were reinterpreted (Fig. S2A and Table S1) and samples with ammonium acetate coprecipitation were analyzed by HPLC-ESI-ToF-MS (Fig. S2B).

Spectra of samples without ammonium acetate were collected on an Exactive Plus EMR Orbitrap spectrometer with an electrospray source at the ThermoFisher's facility in San Jose, CA with sample infusion rate = 5 µL/min; particle concentration ~ 0.5 µg/µL; scan duration = 4.8 minutes (*25*). In the present analysis, the spectra were recalibrated using the observed signals of the z = 6- to 9- ions of (101,42), the dominant fragmentation product of the parent of precise composition (102,44) (*11*). Indeed, the recalibrated spectra in Figure S2A show better agreement with a composition of (146,57). The experimental parent mass of the (146,57) signal is 37,489 Da while the theoretical mass is calculated to be 37,488 Da, an error of 1 Da. Furthermore, one of a series of satellite peaks, previously assigned to an impurity adduct of ~63 Da mass, is now reinterpreted as arising from (144,60) and its ESI fragmentation products (144,58) and (143,58). The experimental mass for (144,60) is 34,551 Da, five Da away from the theoretical value of 37,556. Thus, the reinterpretation of the ~63 Da spacing to the substitution of two gold atoms for three ligands (*2,3*) is consistent with the simultaneous presence of both (144,60) and (146,57).

In order to estimate the $Au_{146}/Au_{144}$ relative abundances in the cluster sample without ammonium acetate coprecipitation based on mass spectrometry, the signals for z = 5-, 6-, 7-, 8- charge states for (146,57) and (144,60) were integrated and compared as shown in Table S1. This exercise indicates that the ratio of (146,57) to (144,60) is ~ 1.8, or nearly double. Of course, this estimate should be taken with caution as the relative signal strengths are biased by how easily each component dissolves, spray, and ionizes.

The cluster samples with ammonium acetate coprecipitation were analyzed by the LC-MS method, which resolves (separates) multi-component samples (chromatographically) prior to the identification of each component (mass-spectrometrically). It can also help in reducing interference from adducts (*42, 43*). The $Au_{144}/Au_{146}$ clusters were detected intact (Fig. 2SB). The minor co-components are separated cleanly, however no resolution of (144,60) from (146,57) was achieved. The major peak could be cleanly attributed to the intact (146,57) species, in the charge-states = [4-] thru [7-], with [5-] dominant. Minor satellite peaks could be attributed to water adduction. The peak signal that could be attributed to intact (144,60) was never stronger than ~ 1/4th of the base peak (146,57). On this basis, it is concluded that the sample has a significantly reduced (144,60) abundance in relation to the dominant (146,57) component.

Computational DFT methods.

The *ab initio* DFT structural optimization started from the experimental structure. In order to obtain a tractable size for the calculations, we have replaced both the sulfur atom and the rest group of all ligands by chlorine atoms (isoelectronic substitution, (*44*)). Subsequently, the structure was relaxed using the VASP code (*45,46*) with PAWs until all forces were smaller than 0.001 eV/ Å. A cubic unit cell of 33 Å and an energy cutoff 280 eV was employed. A charge state of 3- (three additional electrons) was used to comply with the expected electronic shell closing at 2008 electrons according to the superatom model. Among our various optimizations with distinct exchange-correlation functionals and charge states {3+, 1+, 1-, 3-}, the LDA optimization chg.= 3- turns out to best represent the Au-Au bonds (Fig. S13), unlike the calculations done using a GGA functional (PBE; not shown).

Figure S13 shows the Au-Au distances are very well reproduced, which holds both for the nearest-neighbor distances and for Au-Au distances of more distant Au atoms. Unlike the nearest-neighbor Au-Au bonds, the Au-S bond lengths show a bimodal distribution, in agreement with experiment, where this double-peak structure is rather broad. These peaks get much narrower following the relaxation, indicating a slight symmetrization of the structure compared to the irregularities from diffraction. The first peak is at 2.297 Å, the second at 2.360 Å. The minimum separating the two peaks of the distribution is at 2.330 Å. The curves are obtained by convolution with a gaussian of σ= 0.01 Å. Similar results were obtained on the -SCH$_3$ model where the R group is replaced by methyl CH$_3$.

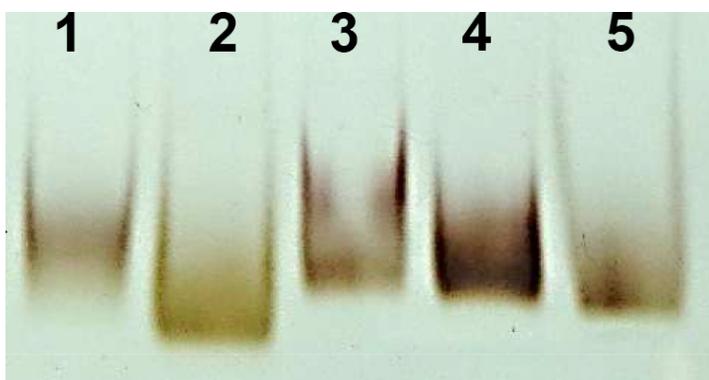

**Fig. S1.**

Gel electrophoresis of gold nanoclusters. (1) Au$_{144}$ or ~29kDa (core mass) standard*. (2) Au$_{102}$ or ~22kDa (core mass) standard*. (3) and (4), two different aliquots of Au$_{146}$ without ammonium acetate coprecipitation. (5) Au$_{146}$ with ammonium acetate coprecipitation. *Samples analyzed by Mass spectrometry revealing peaks traditionally assigned to Au$_{144}$ or Au$_{102}$ (*11, 25*).

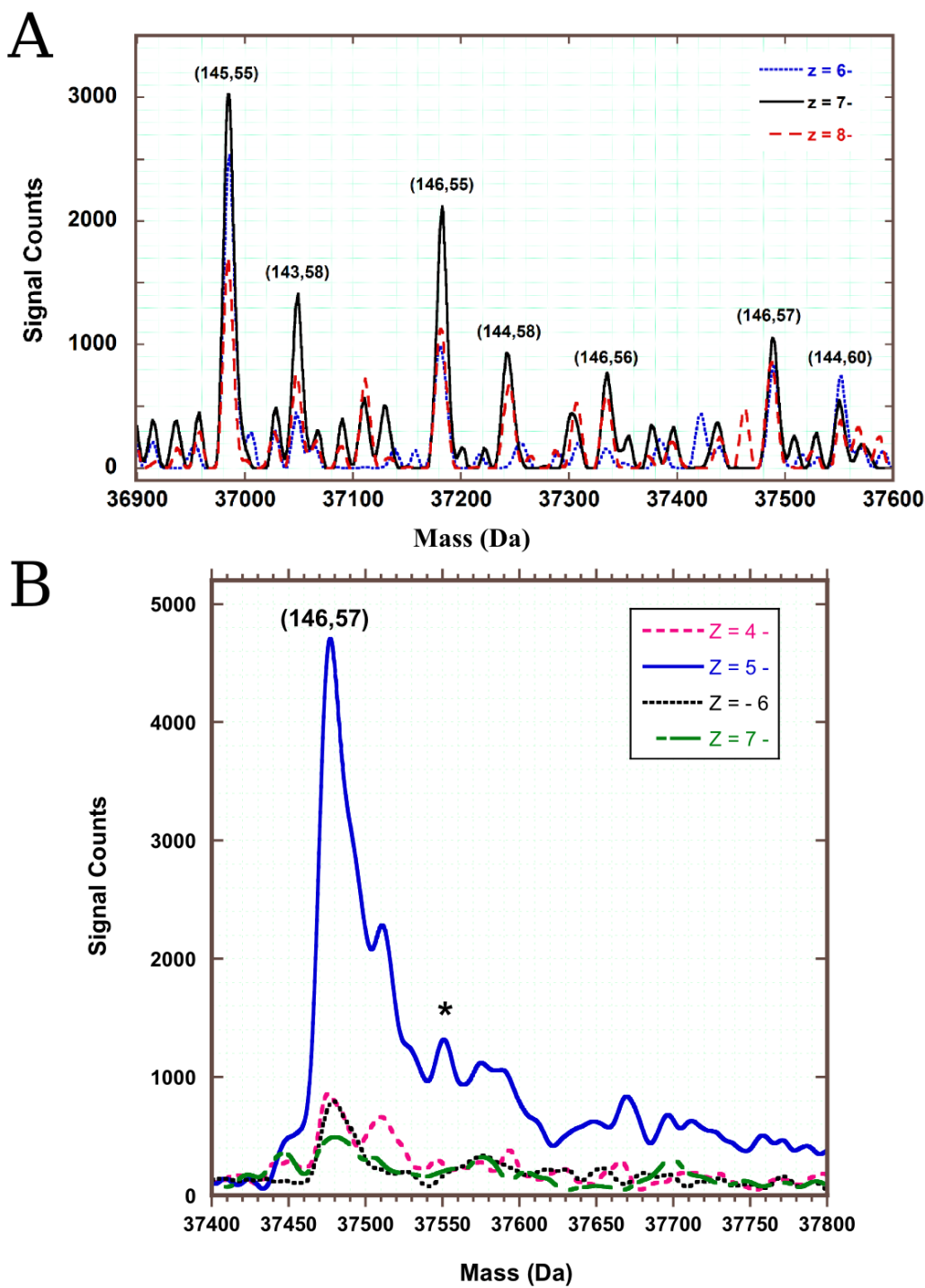

**Fig. S2**

Mass Spectrometry analysis in the mass region (146,57) - (144,60). (**A**) Orbitrap ESI-MS of sample without ammonium acetate coprecipitation. (**B**) HPLC-ESI-ToF-MS of sample with ammonium acetate coprecipitation. The labeled peaks were assigned by multiplying the calibrated ion signals (m/z) by the indicated charge (z = 4-, 5-, 6-…). (*) marks expected position for (144,60).

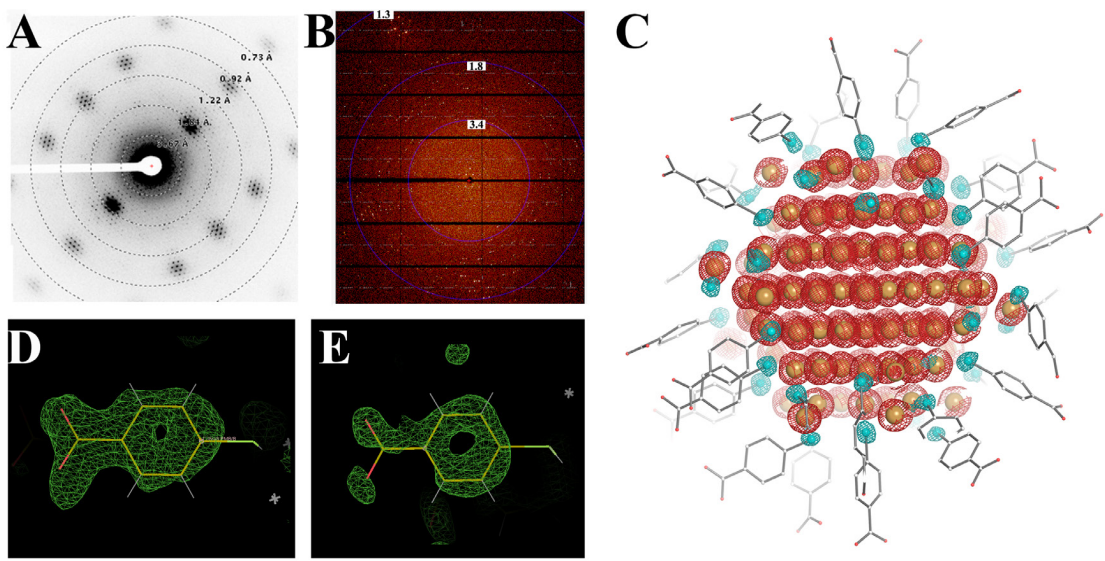

**Fig. S3**

Typical diffraction patterns obtained with (**A**) MicroED at subatomic resolution (outer ring at 0.73Å) and (**B**) X-ray at atomic resolution (outer ring at 1.3 Å). Subatomic resolution could only be achieved with electron diffraction despite multiple attempts at three synchrotron sources (See Table S2) (**C**) Final refined X-ray 2Fo-Fc map contoured at 2 σ from data set 1 (See Table S2). The presence of all 57 p-MBA were only evident in data set 1 (see methods). (**D**) and (**E**) Fo-Fc map (contoured at 2 σ and utilizing +15 B-sharpening factor in Coot (47)) illustrating difference density for p-MBA molecules from data set 1.

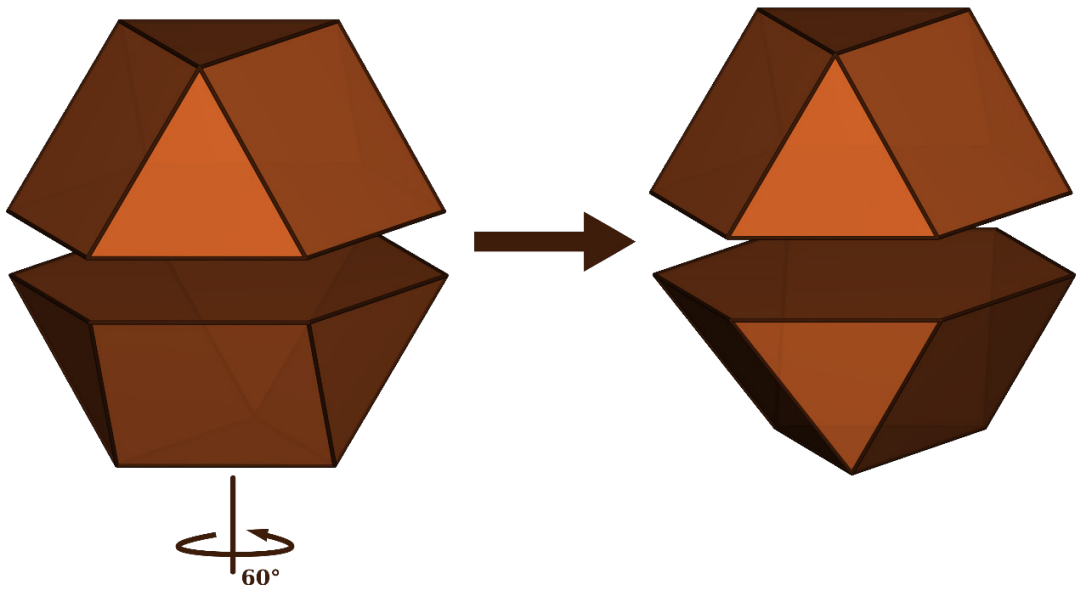

**Fig. S4**

Conversion of cuboctahedron (left) to anticuboctahedron, or J27 (right). A cuboctahedron is the smallest polyhedron to describe the FCC packing whereas nested anticuboctahedra represent a twinned-FCC structure.

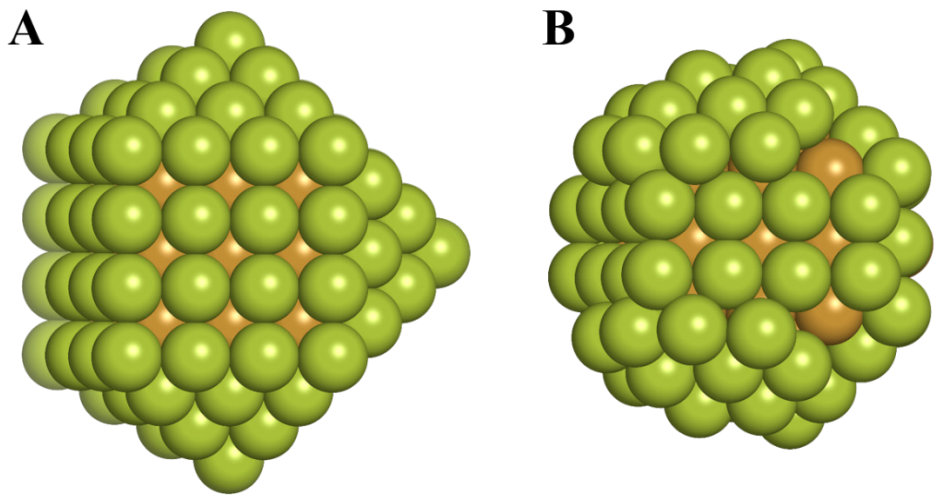

**Fig. S5**
(**A**) Theoretical three-shell J27 and (**B**) incomplete three-shell J27 in $Au_{146}$. J27-3 displayed in green. The truncation of atoms in $Au_{146}$ spherizes the cluster.

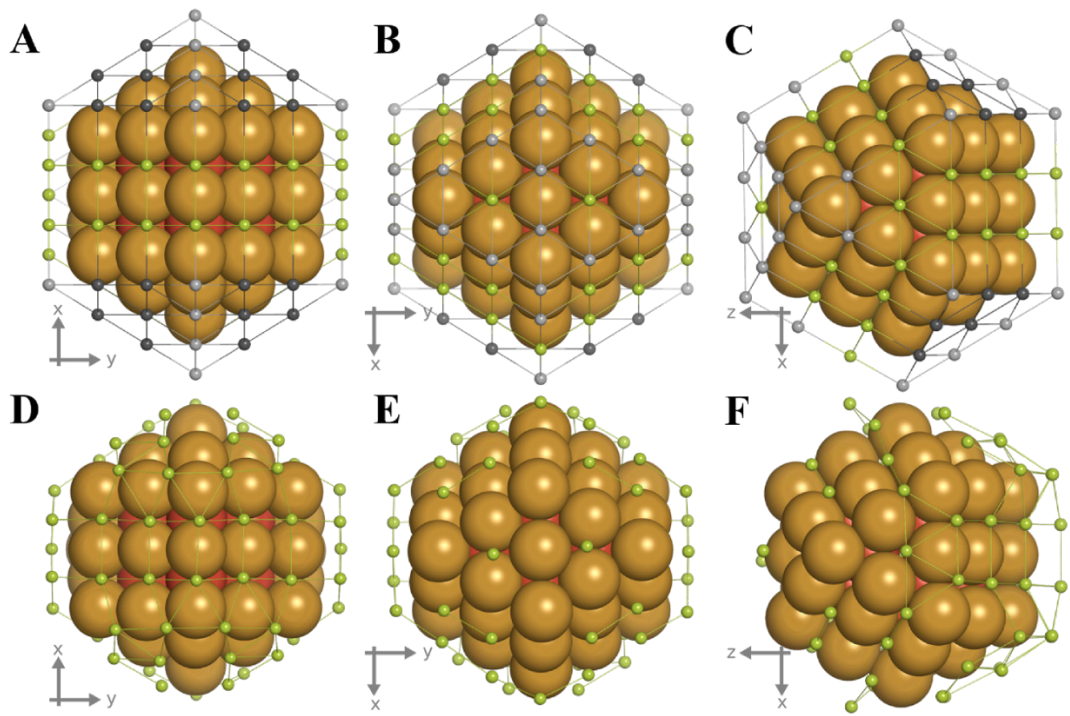

**Fig. S6**

Comparison of the ideal three-shell J27, with 147 atoms, (**A-C**) and the three-shell 115-atom core in Au$_{146}$ (**D-F**). Gold atoms truncated in experimental structure appear in light gray in the ideal structure, while atoms with shifted positions in Au$_{146}$ appear in dark gray.

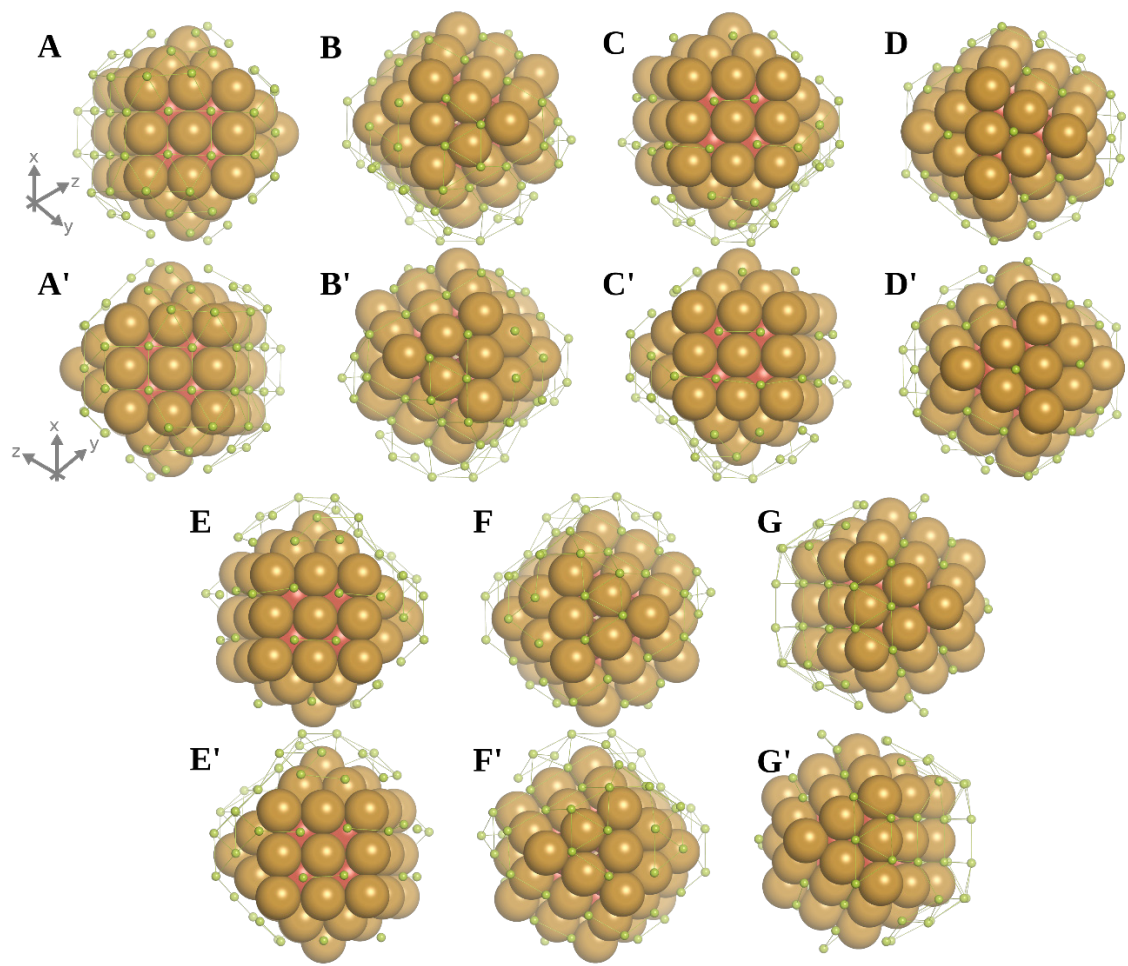

**Fig. S7**

Comparison of mirror planes. Starting with the twin plane parallel to xz plane (view in Fig. S6-D), the structure was rotated to the left (**A**) or the right (**A'**) to obtain views of mirror planes. Sets of mirror planes views (**B-F**, **B'-F'**) were obtained after consecutive 60-degree rotation around y-axis. (**G**) and (**G'**) corresponds to the two mirror planes parallel to the twin plane.

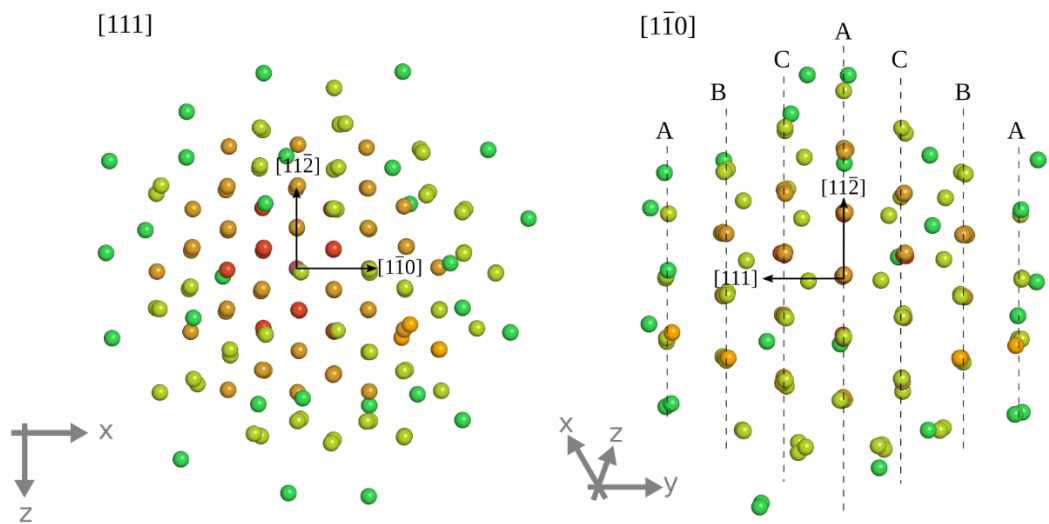

**Fig. S8**

View of crystallographic planes. Plane directions are absolute to an FCC unit cell. First shell J27-1, red; second shell J27-2, brown; third shell J27-3, light green; extra atoms in the core, dark yellow; peripheral atoms in dark green. Most peripheral atoms deviate from crystallographic positions.

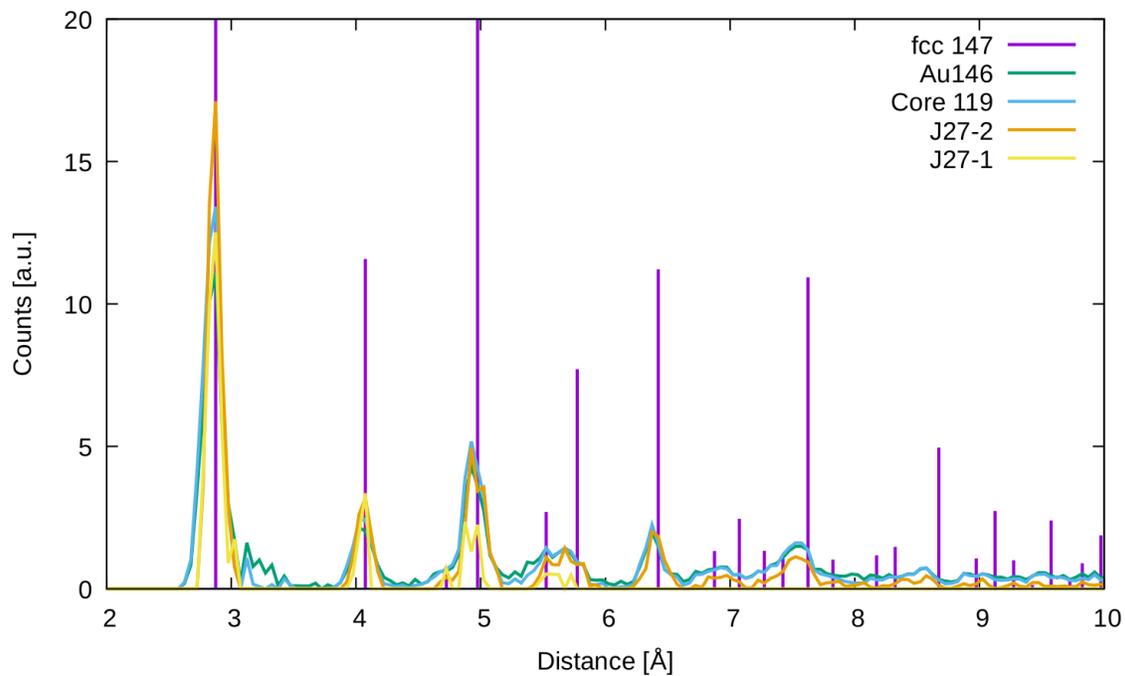

**Fig. S9**

Radial pair distribution function. Line in purple represents ideal atomic distances retrieved from an ideal three-shell FCC structure of 147 gold atoms.

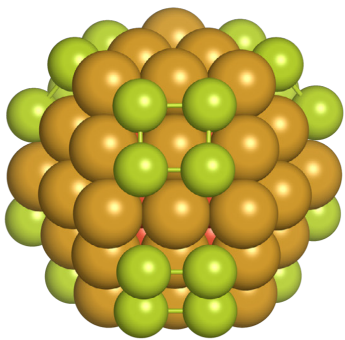 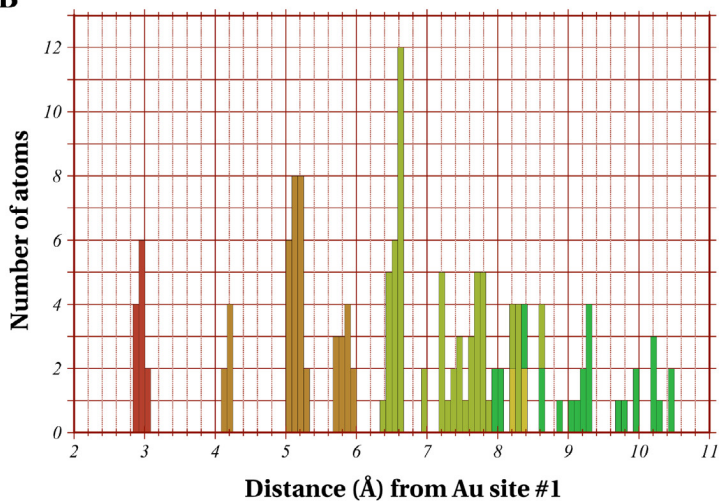

**Fig. S10**
(**A**) 79-atom twinned, truncated octahedron extracted from the experimental core of $Au_{146}$.
(**B**) Radial distance from central atom. It is distinguishable from the histogram the group of 12 atoms (~2.9 Å), corresponding to first shell J27-1, in red; the groups of 6 (~4.2 Å), 24 (~5.1 Å), and 12 (~5.9 Å) atoms forming the second shell J27-2, in brown; and an additional group of 24 atoms (~6.6 Å) belonging to the third shell J27-3, in light green, for a total of 79 atoms. The remaining atoms (distances from central atom > 7 Å) display less ordered positions. J27-1, red; J27-2, brown; J27-3, light green; four extra atoms in the core, dark yellow; peripheral atoms, dark green.

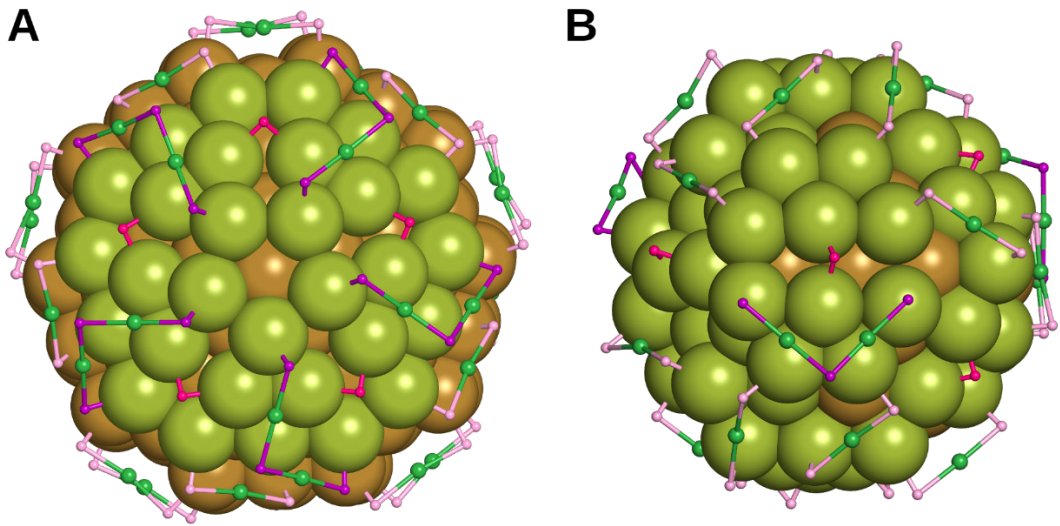

**Fig. S11**

Comparison of dimeric staples (Sulfur in purple) and bridging motifs (sulfur in pink) in (**A**) $Au_{246}$ and (**B**) $Au_{146}$.

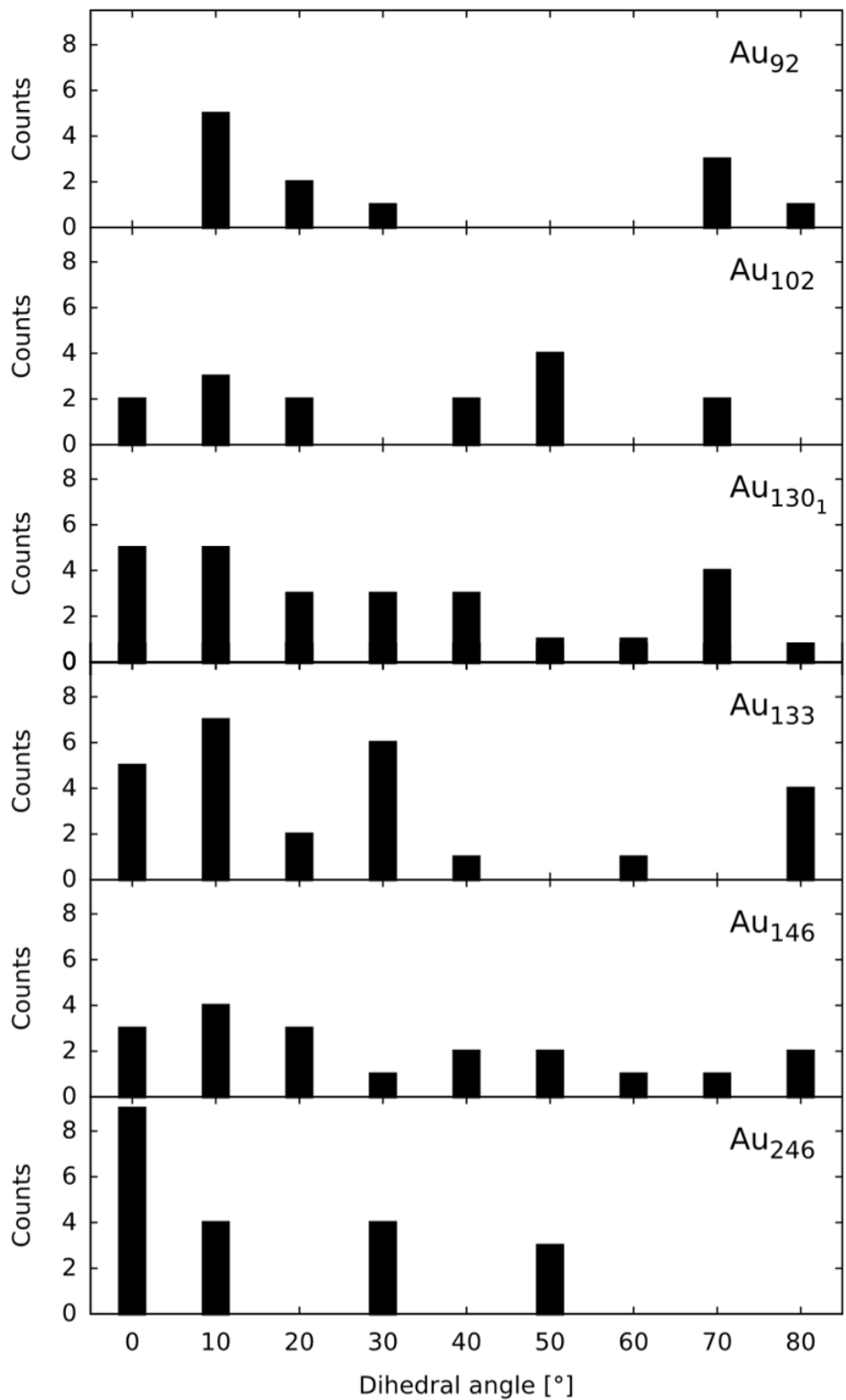

**Fig. S12**

Comparison of dihedral angle in monomeric staples in several clusters.

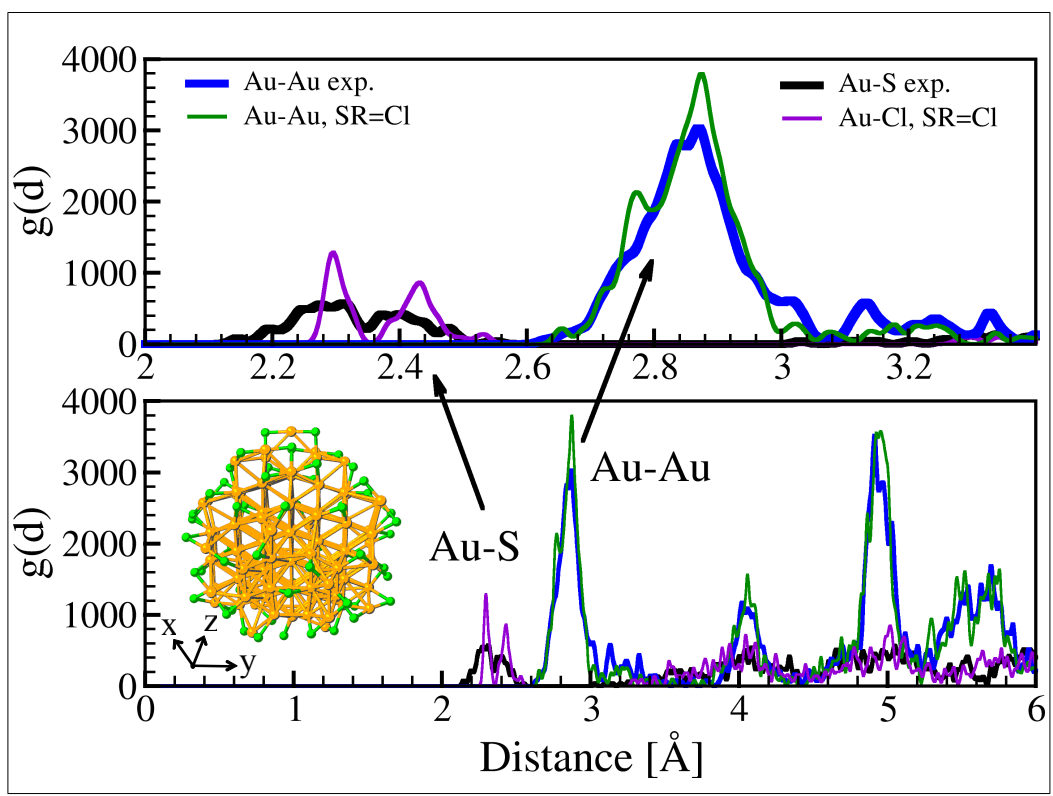

**Fig. S13**
Interatomic distances Au-Au and Au-S bonds derived from experimental (exp.) coordinates and relaxed coordinates from the DFT calculations (Cf. Fig. S9). The Cl-substituted model is shown at the bottom left panel; orientation is shown as indicated by the axes; yellow and green spheres correspond to the Au and Cl atoms, respectively.

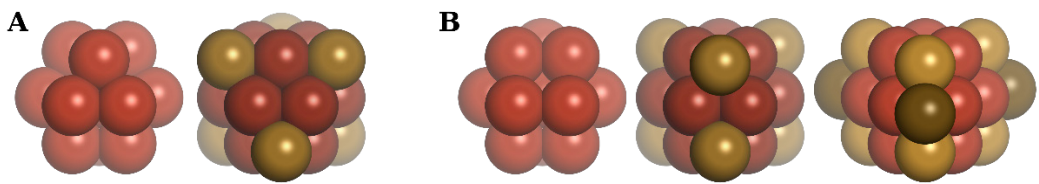

**Fig. S14**

Effect of a twin on the growth of nanoparticles. Growth on (**A**) a cuboctahedron and (**B**) an anticuboctahedron (J27). In both cases, the 13-atom kernel (red) could be first capped with an atom in each of the {100} facet (brown). The remaining available sites in the cuboctahedron are 3-coordinated, resulting in a 19-atom seed. In the J27 the two atoms deposited in mirror {100} planes create 4-coordinated sites on the twin plane, resulting in a 22-atom seed.

**Table S1.**
The signal from the intact (146,57) component is approximately twice that from the intact (144,60) component.

| Charged Species | Au₁₄₆ [5-] | Au₁₄₄ [5-] | Au₁₄₆ [6-] | Au₁₄₄ [6-] | Au₁₄₆ [7-] | Au₁₄₄ [7-] | Au₁₄₆ [8-] | Au₁₄₄ [8-] |
|---|---|---|---|---|---|---|---|---|
| Exp. Parent Mass (Da) | 37487.6 | 37549.2 | 37490.2 | 37552.1 | 37488.3 | 37550.2 | 37488 | 37551.4 |
| Signal Intensity (counts) | 271.8 | 189.2 | 842.8 | 746.8 | 1053.6 | 553.3 | 859.4 | 389.8 |
| FWHM (Da) | 12 | 11.8 | 11.6 | 11.1 | 11.2 | 10.4 | 11.5 | 9.7 |
| Area (a.u.) | 3255 | 2240 | 9755 | 8258 | 11845 | 5779 | 9876 | 3788 |
| Expected (Mass, Da) | 37488.4 | 37553.97 | 37488.4 | 37553.97 | 37488.4 | 37553.97 | 37488.4 | 37553.97 |
| Mass Error (Da) | 0.8 | 4.7 | -1.8 | 1.8 | 0.1 | 3.8 | 0.4 | 2.6 |
| Au₁₄₆ / Au₁₄₄ Ratio (Avg = 1.8) | 1.5 | | 1.2 | | 2 | | 2.6 | |

**Table S2.**
Crystallographic data and structure statistics.

|  | **X-ray Dataset 1** (SLS: XO6SA – PXI) | **X-ray Dataset 2** (APS: SERCAT – 22ID) | **X-ray Dataset 3** (APS: GM/CA – 23IDB) | **microED Dataset** (Janelia: Titan Krios) |
|---|---|---|---|---|
| **Data Collection** | | | | |
| *Wavelength* | 1.05 Å | 1.00 Å | 0.67 Å | 0.025 Å |
| *Reflections* | | | | |
| Measured | 179954 | 125543 | 144128 | 205619 |
| Unique | 58950 | 43620 | 28145 | 56295 |
| *Resolution (Å)* | 32-1.13(1.2-1.1) | 40-1.7(1.6-1.5) | 35-1.4(1.5-1.4) | 12.4-0.87(0.9-0.8) |
| $R_{merge}$ *(%)* | 8.7 (51) | 7.3(47) | 12 (32) | 57.9 (89.8) |
| *I/$\sigma$(I)* | 5.13 (0.74) | 6.3(1.4) | 5.4 (1.2) | 5.27 (0.36) |
| *Completeness (%)* | 84.5 (48.4) | 95.2(83) | 87 (52) | 70.2 (35.3) |
| *Multiplicity* | 2.57 (0.93) | 4.5(2.2) | 2.3 (0.8) | 1.33 (0.2) |
| **Refinement Statistics** | | | | |
| *Space Group* | P12(1)/c | P12(1)/n | P12(1)/n | P12(1)/n |
| *Cell (a, b, c)* | 56.2, 32.5, 57.4 | 31.7, 56.2, 51.5 | 31.7 56.1, 51.4 | 31.7, 56.2, 51.5 |
| *Angles ($\alpha$, $\beta$, $\gamma$)* | 90, 116.6, 90 | 90, 90.5, 90 | 90, 90.5 90 | 90, 90.5, 90 |
| $R_{work}/R_{free}$ | 19/21 | 18/21 | 22/24 | 27/29 |
| *Number of Atoms* | | | | |
| Au | 146 | | | |
| S | 57 | | | |
| *Number of Molecules* | | | | |
| p-MBA | 57 | | | |

**Movie S1**

360-degree rotation of $Au_{146}$(p-MBA)$_{57}$ with the twin plane in the horizontal plane. First shell J27-1, red; second shell J27-2, brown; third shell J27-3, light green; surface or peripheral gold atoms, dark green; sulfur, cyan; carbon, light gray, and oxygen, light red.